# Spectral measurement of the breakdown limit of β-Ga$_2$O$_3$ and tunnel ionization of self-trapped excitons and holes

Md Mohsinur Rahman Adnan[1,&], Darpan Verma[2,&], Zhanbo Xia[1], Nidhin Kumar Kalarickal[1], Siddharth Rajan[1], Roberto C. Myers[1,2]

[1]Department of Electrical and Computer Engineering, The Ohio State University, Columbus, Ohio 43210, United States

[2]Department of Material Science and Engineering, The Ohio State University, Columbus, Ohio 43210, United States

Abstract:

**Owing to its strong ionic character coupled with a light electron effective mass, β-Ga$_2$O$_3$ is an unusual semiconductor where large electric fields (~1-6 MV/cm) can be applied while still maintaining a dominant excitonic absorption peak below its ultra-wide bandgap ($E_g$~ 4.6- 4.99 eV). This provides a rare opportunity in the solid-state to examine exciton and carrier self-trapping dynamics in the strong-field limit at steady-state. Under sub-bandgap photon excitation, we observe a field-induced red-shift of the spectral photocurrent peak associated with exciton absorption and threshold-like increase in peak amplitude at high-field associated with self-trapped hole ionization. The field-dependent spectral response is quantitatively fit with an eXciton-modified Franz-Keldysh (XFK) effect model, which includes the electric-field dependent exciton binding energy due to the quadratic Stark effect. A saturation of the spectral red-shift with reverse bias is observed exactly at the onset of dielectric breakdown providing a spectral means to detect and quantify the local electric field and dielectric breakdown behavior in β-Ga$_2$O$_3$. Additionally, the field-dependent responsivity provides insight to the photocurrent production pathway revealing the photocurrent contributions of self-trapped excitons (STXs) and self-trapped holes (STHs) in β-Ga$_2$O$_3$. Photocurrent and p-type transport in β-Ga$_2$O$_3$ are quantitatively explained by field-dependent tunnel ionization of excitons and self-trapped holes. We employ a quantum mechanical model of the field-dependent tunnel ionization of STX and STH in β-Ga$_2$O$_3$ to model the non-linear field-dependence of the photocurrent amplitude. Fitting to the data, we estimate an effective mass of valence band holes (18.8 m$_0$) and an ultrafast self-trapping time of holes (0.045 fs). This indicates that minority-hole transport in β-Ga$_2$O$_3$ can only arise through tunnel ionization of STH under strong fields.**

[&]These authors contributed equally to the work.

# I. INTRODUCTION

The conversion of electromagnetic radiation (photon flux) into current density (charge flux) is of fundamental importance in all solid-state optoelectronics. An electron-hole pair is produced at the location of single photon absorption and therefore the mutual Coulombic bond between the photocarriers (exciton) must be broken (dissociation) before carrier separation and collection can take place. However, excitons are generally considered unimportant for understanding the room-temperature optoelectronic response in dominantly covalent (sp$^3$ bonded) crystalline semiconductors [1,2]. Traditionally called Wannier-Mott excitons [3,4], the small exciton binding energies ($E_X$ < 25 meV) and small dielectric constants (K) combined with light effective masses ($m^*$<1) of the host material lead to exciton Bohr radii typically spanning multiple nearest neighbor bond distances ($a_B$ > 1 nm) and are assumed to be fully thermally or electrically dissociated in room temperature optoelectronics [2,5,6]. The opposite extreme occurs in strongly ionic solids where exciton absorption dominates the optical response with $E_X$ > 500 meV and $a_B$ < 0.5 nm [7,8]. However, ionic crystals with such strongly-bonded excitons can hardly be classified as semiconductors, e.g. alkali halides [9,10], with strong electron-lattice coupling strongly limiting charge transport (polarons) [11,12].

More recently, the emergence of two-dimensional (2D) semiconductors with sizable bandgaps and light masses, e.g. transition metal dichalcogenides (TMDs), has demonstrated a regime of exciton physics with simultaneously large $E_X$ and $a_B$ [13–16]. Under these conditions, room temperature optoelectronic properties are dominated by excitons near the band edge. Moreover, charged excitons known as trions become possible with interesting physics and potential optoelectronic applications [17–21]. Compared to these TMDs, the ultra-wide bandgap (~4.6-4.99 eV [22,23]) semiconductor β-Ga$_2$O$_3$ exhibits similar exciton properties to the 2D materials, but with the added tunability associated with its high dielectric breakdown field limit (~6-8 MV/cm [22–24]) motivating its development for high-power high-field devices and solar-blind ultraviolet photodetectors [23,25–28]. Additionally, these properties enable exploration of steady-state exciton dynamics in the strong-field (non-perturbative) limit, Fig. 1a and b.

## A. eXciton Franz-Keldysh (XFK) Effect

Band-to-band absorption is altered in the application of an electric field, characterized by a broadening of the absorption edge below the bandgap due to tunneling-assisted absorption, called the Franz-Keldysh (FK) effect [29,30], Fig. 1c. In 1966, Aspnes [31] developed a quantitative model for the field-dependent absorption spectrum within the effective mass approximation. This model was recently applied by Maeda *et al.* to quantitatively model the absorption phenomenon at constant excitation wavelength with varying applied field in wide bandgap semiconductors like SiC and GaN based devices [32–35]. Sub-bandgap absorption in GaN gives rise to excitons which must be dissociated before any photocurrent can be measured [36], but the Aspnes model ignores exciton absorption and is therefore only valid in GaN in a limited range of wavelengths and fields. Theoretical work by Dow and Redfield [37], Blossey [38], and Merkulov [39] showed that the electron-hole Coulombic bond (exciton) leads to a qualitatively distinct electro-absorption behavior from the FK effect, namely a sub-bandgap absorption peak that shifts under an applied field, which we refer to as the eXciton Franz Keldysh (XFK) effect, Fig. 1c. Exciton absorption was observed in the FK-effect in GaN [40,41], revealing an exciton peak structure near the band edge in GaN absorption spectra, which redshifts with increasing applied field. We recently reported the experimental observation and quantitative modeling of the XFK effect in the photocurrent spectral response of GaN [36], which demonstrates a spectral measurement of the local electric field and a route to develop all-optical electric field microscopy to explore breakdown physics and refine electrostatic modeling for high field devices. However, in the case of GaN, $E_X$ ~20.3-27 meV [36], is modest (~$k_B T$), and therefore the XFK peak is only resolvable at room temperature at relatively low electric fields since the excitons are easily dissociated [40,41]. By contrast, excitons in TMDs and β-$Ga_2O_3$ fully dominate the sub-bandgap absorption, an effect accentuated in the latter by the weak band-to-band absorption of its indirect bandgap [42–44].

## B. Similarity of Excitons in 2D TMDs and β-$Ga_2O_3$

Because excitons are electrically neutral, photocurrent is only obtained from strongly bound excitons if a sufficiently large electric field is applied to dissociate them into mobile conduction band electrons and valence band holes via tunnel ionization [45–49], illustrated in Fig. 1a, otherwise the excitons would decay through electron-hole recombination. The tunneling barrier height and width are modified by an expected increase in the binding energy with field due to the Stark effect [13–15,40,41,50–52]. Such field ionization phenomena are well studied in organic semiconductors [52,53], but for inorganic solid-state systems were typically limited to discussion of impurity ionization in the low-field (perturbative) regime [40,41,49]. Only very recently have such high-field exciton-dissociation processes been described in a group of inorganic solid-state systems i.e. in TMD materials [13–15,50,51]. For example, in the layered semiconductor $WSe_2$ with $E_X$ = 170 meV, the bias-dependent photocurrent amplitude was explained in terms of a Stark-modified exciton binding energy, see supplementary material of Ref. [14]. *Ab initio* Bethe-Salpeter equation (BSE) calculations could predict the bandgap correctly only if dielectric screening effects of several h-BN layers were included but failed to predict the correct $E_X$, while a screened Wannier-Mott model could explain the observed $E_X$. Previously, numerical and analytical models for quantitative field-induced dissociation rates for different TMDs showed the dependence of exciton dissociation on the anisotropic dielectric screening environment [17,50,54–57], where the in-plane electron-hole potential follows a 1/r potential over long range, but a weaker logarithmic divergence over short range due to screening [37,38]. More recently, Kamban and Pedersen treated excitons as 2D Hydrogen atoms confined in-plane within a given screening distance to model exciton dissociation and Stark effects in TMDs [50].

Similar 2D-like Hydrogenic excitons appear to be present in $β-Ga_2O_3$. The optical anisotropy of $β-Ga_2O_3$ was demonstrated experimentally by Onuma *et al.*'s electro-reflectance study [43] and further *ab initio* calculations (DFT+GW) by Furthmüller *et al.* [44] showing that $O^{2-}$ 2p valance band states (holes) of $β-Ga_2O_3$ couple with $Ga^{3+}$ 4s conduction band states (electrons) forming anisotropic (2D-like) excitons; absorption energies for $s - p_x$ (a-axis polarized) and $s - p_z$ (c-axis polarized) are ~0.3 eV larger than along $s - p_y$ (b-axis polarized). Exciton wave functions in $β-Ga_2O_3$ therefore exhibit a 2D anisotropy; weakly bound with large wave functions along the b-axis [010], and tightly bound with small wave function within the ac-plane (010). This pseudo-2D behavior of excitons in $β-Ga_2O_3$ is further corroborated by BSE calculations of Varley and Schleife [42] showing a strong exciton absorption at lower energy for photons polarized along a (x) or c (z) axis, compared with b (y) axis. Just like TMDs, a screened Mott-Wannier model for estimating the correct $E_X$ = 270 meV must be used for $β-Ga_2O_3$ [60]. Beschedt *et al.* showed that due to the strong exciton binding energy (larger than optical phonon energies), lattice screening of the electron-hole interaction is greatly reduced in comparison with what is expected from the static dielectric constant. Similar to TMDs, β-Ga$_2$O$_3$ exhibits a large E$_X$ (~189-270 meV [60–62]) and anisotropic (2D-like) optical response as predicted from the BSE calculations of its absorption spectra by Varley and Schleife [42]. Due to the anisotropic dielectric environment, similarly large exciton polarizabilities (induced dipole) are expected [50] such that the exciton binding energy is strongly field dependent (Stark effect).

### C. Exciton Stark effect and tunnel ionization of self-trapped holes in β-Ga$_2$O$_3$

Similar to polarons ubiquitous to many ionic materials [63], β-Ga$_2$O$_3$ exhibits carrier self-trapping, where due to the large effective mass of the holes, they are localized to a few bond distances leading to a strong lattice distortion. A valence band hole centered an O$^{2-}$ site will push Ga$^{3+}$ nearest neighbors outward and draw O$^{2-}$ next nearest neighbors inwards, forming a trapping potential (~490 meV [64]) that then localizes the hole even further. These self-trapped holes (STHs) are thought to limit the p-type conductivity in β-Ga$_2$O$_3$. Under optical absorption, the STHs together with free electrons form Self-Trapped eXcitons (STXs) which demonstrate high binding energy (~680 meV [64]) similar to alkali halide excitons. Yamaoka *et al.* observed STX and STH in the optical response of β-Ga$_2$O$_3$ [64–66]. These STXs and STHs must be field-ionized if photocurrent is to be obtained from β-Ga$_2$O$_3$ devices, Fig. 1a.

Here, we present measurements of β-Ga$_2$O$_3$ exciton absorption and dissociation processes in the strong-field regime. A photocurrent spectral peak associated with exciton absorption red shifts under reverse bias due to the XFK effect. The spectral response is explained by including the quadratic field-dependence of the exciton binding energy, the exciton Stark effect. The model predicts a continuous red shift of the exciton peak with field, however the data reveal a saturation effect, where the spectral red shift halts due the onset of breakdown, in agreement with previously reported estimates for the breakdown field (~6 MV/cm [23,24]). Using the spectral peak position as an electric field sensor, we examine the photocurrent peak amplitude response as a function of field. A high-field threshold behavior is observed, with a turn-on at ~4.5 MV/cm. The peak amplitude changes are interpreted as due to the field-dependent quantum efficiency (ratio of photocarriers collected to photons absorbed) associated with the exciton and STH ionization processes. We adapt a field-ionization model of quasi-bound state tunneling to describe the exciton dissociation process in β-Ga₂O₃. Using literature values of minority carrier lifetimes and carrier mobilities we derive field-dependent rate equations for dissociation, ionization, and photo-carrier transport. The non-linear field-dependence of the photocurrent is fit using two free parameters: the valence band hole effective mass, and the self-trapping time of holes. The determined values agree with theoretical band structure estimates for the valence band effective mass [60,61] and recent ultrafast measurements of self-trapped holes [67].

## II. FIELD-INDUCED RED-SHIFT OF EXCITON ABSORPTION PEAK:

Figure 1b shows the structure of β-Ga₂O₃ Schottky Barrier Diode (SBD) used for our study. Epitaxial growth of β-Ga₂O₃ is carried out using PAMBE (Riber MBE Control Solutions M7). Slightly Ga-rich growth conditions are used for the growth of β-Ga₂O₃: at a Ga beam equivalent pressure (BEP) of 8×10⁻⁸ Torr, and oxygen pressure during growth of 1.5x10⁻⁵ Torr, plasma power of 300 W and $T_{sub}$= 630 °C (pyrometer). The epitaxial structure for the SBD consists of a UID Ga₂O₃ layer of $1\ \mu m$ thickness grown on a Sn: Ga₂O₃ (010) substrate [22,68] (Tamura). The Schottky top contacts are formed by evaporating a Ni/Au (30 nm/100 nm) metal stack while the ohmic bottom contacts are a Ti/Au (50 nm/100 nm) metal stack annealed at 470 °C for 1 min. As shown in Fig. 1 band diagram, the UID Ga₂O₃ layer has an average doping concentration of $5 \times cm^{-3}$, obtained from C-V measurements. Beyond the UID layer, the doping concentration is slightly higher i.e. $1 \times 10^{18}\ cm^{-3}$. The energy band diagram calculated from 1-D Poisson solver [69,70] in Fig. 1b demonstrates that almost all the applied potential drops within the $1\ \mu m$-thick UID layer. The experimental setup for the photocurrent spectral measurement is the same as described in Ref. [36]. A chopped Xe-lamp coupled to a monochromator serves as the wavealength tunable- pulsed- photoexcitation that is focused using a reflective microscope objective (40×) onto the SBD generating photocurrent, which is acquired with a lock-in amplifier.

The device and experimental geometry are shown in Fig. 2a, where the crystallographic axes of the monoclinic β-Ga₂O₃ crystal are shown as well as the two positions (P1 and P2) of local photocurrent spectroscopy. Representative photoresponsivity spectra for β-Ga₂O₃ (at P1) are shown in Fig. 2b and 2c at various values of reverse bias. The spectral variation in the absorption due to applied reverse bias ($V_{exp}$) is qualitatively different for the β-Ga₂O₃ device than what is observed in other wide bandgap semiconductors [33,36,40,41]. GaN shows an XFK induced redshift of the optical absorption edge along with spectral broadening as bias increases [36]. This increased sub-band-gap absorption broadening is typical of the band edge FK-effect. In contrast, β-Ga₂O₃ shows an XFK-like red shift of the eXciton peak with increasing bias (Fig. 2b), but without any apparent spectral broadening (Fig. 2c). Additionally, unlike most semiconductors the photocurrent does not increase strongly above bandgap (4.99 eV), instead showing reduced absorption. This spectral behavior is consistent with the literature on excitonic absorption examined both theoretically and experimentally. The characteristic exciton peak of β-Ga₂O₃ has been predicted by solutions of Bethe Spelter Equation (BSE) [42]. The ab-initio calculation of imaginary part of dielectric function of β-Ga₂O₃ has been carried out in [42] both with and without accounting for the e-h pair interaction. Without the exciton effect, the absorption response is a slow increase due to the indirect bandgap. Only when the e-h interaction is accounted for, a strong excitonic peak is observed below bandgap. The excitonic peaks were also observed from spectroscopic ellipsometry measurements in the imaginary part of the dielectric function by Sturm et al. [62].

As described in Sec. IB, one of the signature features of exciton absorption in β-Ga₂O₃, is the anisotropic absorption. Therefore, to further confirm the excitonic origin of the photocurrent peak, we utilize a linear polarizer (α-BBO), connected to a motorized rotation stage, to measure the polarization dependence of the photocurrent peak position. X-ray diffraction was first used to determine the in-plane crystallographic orientation of the device. Photocurrent spectra are collected at a reverse bias of 30 V with the polarizer angle ($\theta$) stepped every 5° for 360°, where $\theta = 0$, is aligned to the a-axis [100] direction. The peak position of the photocurrent spectra ($E_{ph}^0$) are extracted by fitting the peaks to a bi-Gaussian function. The peak red shift ($E_{ph}^0(0) - E_{ph}^0(\theta)$) relative to the a-axis position is plotted in polar coordinates as a function of polarizer angle ($\theta$) in Fig. 2d, where the high symmetry crystal axes are labeled. As expected, for this (010) plane, in β-Ga$_2$O$_3$ the exciton absorption peak shows 2-fold rotational symmetry. Within the ac-plane (010) the a-axis orientation shows the highest energy absorption. The data show the maximum red shift along the [001] and [102] axes reproducing the anisotropic exciton absorption in β-Ga$_2$O$_3$ previously reported by both experiment and theory [42–44],

From Fig. 2b we observe that the responsivity peak amplitudes, ($I_{PR}^0$) are smaller than reported in previous works [27,28,71]. In these studies, the amplitudes are larger than what is allowed by the theoretical limit of photocurrent generation due to a proposed gain mechanism, where the Schottky barrier gets lowered by positive charge (self-trapped hole) build up at the metal/semiconductor interface [27,28]. In these studies, UV illumination was modulated at a slow rate <1 Hz, and the unusually large responsivities appear to charge-up over >100 ms timescales corresponding to interface charge build-up. In our experiments, we modulate the excitation at ~200 Hz to eliminate these slow dynamics. To confirm that this is indeed the case, we carried out photocurrent spectroscopy as a function of illumination time. Fig. 2e shows the peak amplitude ($I_{PR}^0$) dependence on illumination time, $\frac{1}{2\nu}$ (s), where $\nu$ is the chopper frequency. In Ref. [27], hole trapping and positive charge build up happens only when the junction is under UV illumination for an extended period of time >100 ms) [72]. As can be observed from Fig. 2e, the responsivity increases strongly for timescales approaching 100 ms, whereas the gain mechanism can be ignored for illumination times <5 ms.

To examine the field dependence of the photocurrent spectra, bias dependent measurements were performed while focusing the light source to two different spots on the device: at the edge of the top electrode (P1), and at a displacement along the c-axis from the top electrode (P2); with the excitation polarized along c-axis (Fig. 2a). Since the reverse bias is applied at the top electrode, the potential drop and field variation are expected to be larger at P1 compared to P2. $E_{ph}^0$ and $I_{PR}^0$ are plotted as a function of $V_{exp}$ for data taken at P1 (red) and data taken at P2 (blue) in Fig. 2f and g, respectively. At both positions, we observe a red shift of $E_{ph}^0$ (Fig. 2f) with no qualitative difference in the behavior for the two positions within the noise of the peak fitting. P1 (red, near electrode) has a larger range of variation in peak position than P2 (blue, far from electrode), indicative of a higher local $F_{max}$. The $I_{PR}^0$ shows a non-linear increase with bias (Fig. 2g).

### A. Photoresponsivity and field-dependent absorption coefficient

To understand the photoresponsivity behavior, we note that the SBD is a vertical device with probes placed on the top and bottom metal electrodes. If the photon flux incident on the top electrode is $\varphi_0$, then the flux entering β-Ga₂O₃ is $(1-r)\varphi_0$, where $r$ is the reflectance of the top electrode/semiconductor layer. After absorption in the depletion region, the photon flux exiting the depletion region and entering the non-depleted β-Ga₂O₃ is $(1-r)\varphi_0 e^{-\int_0^W \alpha_{XFK}(F(z),\omega)dz}$, where $\alpha_{XFK}(F(z), E_{ph})$ is the field-dependent eXciton Franz Keldysh (XFK) absorption coefficient, $F(z)$ is the vertical component of the electric field versus depth (z), and $W$ is the depletion width. Since the non-depleted layer has negligible electric field, photocarriers can only be produced and collected within the minority hole diffusion length, $L_n$, with an absorption factor of $e^{-\alpha_n L_n}$, where $\alpha_n$ is the absorption coefficient of n-doped β-Ga₂O₃. Because the absorption coefficient of β-Ga₂O₃ is quite small below the band-gap ($< 10^3\ cm^{-1}$) [73] and the minority hole diffusion length $L_n$ is only ~200-300 nm [74], then $e^{-\alpha_n L_n} \sim 1.000$ and we can fully neglect photocurrent produced in this region. Absorption in the non-depleted n-type β-Ga₂O₃ layer of ~500 µm reduces the photon flux incident on the back surface to $\sim e^{-\alpha_n L} > e^{-100\ *.05}$, which together with the diffuse hemispherical reflection off the back surface of the substrate allows us to ignore absorption due to a second optical pass through the depletion region. Within the depletion region of the Schottky diode where almost all the voltage drops, the absorbed photon flux (converted into excitons) is $(1-r)\varphi_0 \left(1 - e^{-\int_0^W \alpha_{XFK}(F(z),\omega)dz}\right)$. Thus, the photocurrent density generated in the SBD will be,

$$I_{PC} = \eta q(1-r)\varphi_0 \left(1 - e^{-\int_0^W \alpha_{XFK}(F(z),\omega)dz}\right) \qquad \ldots(1),$$

where η is the internal quantum efficiency (IQE) representing the ratio of carriers collected to photons absorbed and q is the electron charge [72]. Normalizing the photocurrent with respect to the measured wavelength dependent input power density ($E_{ph}\varphi_0$) gives the photoresponsivity, $I_{PR}$ (A/W),

$$I_{PR} = \frac{I_{PC}}{E_{ph}\varphi_0} = \frac{\eta q(1-r)\left(1-e^{-\int_0^W \alpha_{XFK}(F(z),\omega)dz}\right)}{E_{ph}} \qquad \ldots(2)$$

The $r$ spectrum of the SBD device is measured over the range $E_{ph}$ = 3.6 eV to 5 eV and found to be roughly constant, $r = 0.15 \pm 0.04$. In modeling $I_{PR}$, we therefore assume a constant value of the pre-factor, (1-r) = 0.85 [72]. Therefore, it has no impact on the $E_{ph}$ dependence of $I_{PR}$, which depends only on the average absorption coefficient in the depletion layer, the depletion width, and the quantum efficiency.

### B. Stark-modified eXction Franz-Keldysh (XFK) effect in β-Ga$_2$O$_3$

β-Ga$_2$O$_3$ has an exciton binding energy $E_x^0$ = 180-270 meV [60–62] at zero applied field (F=0). The binding energy increases with electric field due to the Stark effect, $E_x = E_x^0 + bF^2$, where $b = \frac{9q^2 a_B^2}{8E_x^0}$ is the polarizability [52]. Here, $a_B = a_0 K m_0/\mu_{ex}$ is the exciton Bohr radius, $a_0$ (~53 pm) is the Bohr radius of Hydrogen and $E_x^0 = 13.6\mu_{ex}/(m_0 K^2)$, where $\mu_{ex}$ is the exciton reduced mass and K is the static dielectric constant, which strongly influences the polarizability $b$. Bechstedt *et al.* developed a modified-Wannier-Mott exciton model for β-Ga$_2$O$_3$ which incorporates an effective dielectric constant (K*~3.99) that considers the LO-phonon charge carrier screening at optical frequencies and the anisotropic effective mass [60]. This modified-Wannier-Mott exciton model captures the experimentally determined values of $E_x$ of β-Ga$_2$O$_3$. K (K*) is the static (effective) dielectric constant which equals 14.82 (3.99) [60], $\mu_{ex} = 0.3m_0$ [60]. Using $a_B$ ~ 0.70 nm, $E_x^0$ ~250 meV we calculate, $b = 2.1 \times 10^{-18}$ $eV\left(\frac{m}{V}\right)^2$, which is of the same order of magnitude as for several of the TMDs, see Table II of Ref. [50]. Note that the predicted Stark shift of β-Ga$_2$O$_3$ traverses a similar range as found in TMD materials for the range of applied fields (≤ 0.5 MV/cm) of Ref. [50].

The field-dependent exciton absorption coefficient, $\alpha_{XFK}(F(z),\omega)$, is given by Merkulov [39],

$$\alpha_{XFK}(F(z),\omega) = Cx/\pi^2(\delta^2 x^2 + 1), \qquad \ldots(3)$$

with $x = \frac{8}{f} e^{\left\{-\frac{4\Delta^{\frac{3}{2}}}{3f} - \frac{2}{\sqrt{\Delta}} \ln\left(\frac{8\Delta^{\frac{3}{2}}}{f}\right)\right\}}$, $\delta = \Delta - 1 - \frac{9f^2}{2}$, $f = \frac{qF(z)a_B}{E_x}$, $\Delta = \frac{E_g - \hbar\omega}{E_x}$

where C is a normalization constant which is the same as for the free carrier case [75], such that it does not affect the peak shift. Employing this quantitative model, we utilize $E_g = 4.99$ eV [76], $E_x^0 = 270$ meV [61], $\mu_{ex} = 0.3\, m_0$ [60] with $a_B \cong 0.70$ nm and K* = 3.99 [60].

### C. Spectral observation of the breakdown limit of β-Ga₂O₃

The electric field profile is well approximated by a triangle, typical for Schottky diodes,

$$F(z) = F_{max}\left(\frac{W-z}{W}\right) \quad \ldots(4),$$

where, W (in $\mu m$) is the depletion width given by $W = \frac{2(V_{exp} + V_{bi})}{F_{max}}$, $V_{exp}$ is the applied reverse bias, $V_{bi} = \varphi_B - \frac{KT}{q}\log\left(\frac{N_c}{N_d}\right)$ is the built-in potential barrier of the Ni/Au/β-Ga₂O₃ SBD, $\varphi_B = 1.23\, V$ [77] is the barrier height, $N_c = 3.72 \times 10^{18}\, cm^{-3}$ [78] is the effective density of states for the conduction band, $N_d = 5 \times 10^{17}\, cm^{-3}$ is the UID layer donor concentration, and $F_{max}$ is the field maximum at the Schottky junction. Due to electrostatic variation, it is possible that the field profile changes from the idealized picture, leading to varying $F_{max}$ and/or $V_{exp}$, thus we explore the spectral sensitivity of $I_{PR}$ to such variations in the field profile. First, holding $V_{exp}$ constant, $F_{max}$ is varied to change the field description $F(z)$ as shown in Fig. 3a. Second, keeping $F_{max}$ constant, $V_{exp}$ is varied, Fig. 3b. For all field profiles, the depletion width is determined by the electrostatic requirement that $V_{exp} = \int_0^W F(z)dz$.

Using Eqns. (2-4) the $I_{PR}$ spectra are calculated assuming $\eta = 1$ and plotted in Figs. 3c and d for the two different electrostatic conditions shown in Fig. 3a and b, respectively. First, we note that the Stark-modified XFK model of $I_{PR}$ shows the peak-shaped spectrum for sub-bandgap energies, in agreement with the experimental data of β-Ga₂O₃ (Fig. 2a). In the case of varying $F_{max}$ and constant $V_{exp}$ (Fig. 3a), the modeled $I_{PR}$ spectra in Fig. 3c show a redshift of the $I_{PR}$ peak with $F_{max}$. For the $F(z)$ profile of Fig. 3(b), where the $F_{max}$ is constant (2 MV/cm) but $V_{exp}$ is adjusted from 10 V to 40 V, the $I_{PR}$ spectra show no clear variation (Fig. 3d). Thus, the peak positions of the $I_{PR}$ spectra ($E_{ph}^0$) depend solely on $F_{max}$ and exhibit no dependence on $V_{exp}$. This demonstrates that for β-Ga$_2$O$_3$ SBDs, the photocurrent spectra are apparently insensitive to absorption in the low field regions.

From the modeled spectra, we numerically determine the relationship between $F_{max}$ and $E_{ph}^0$ in order to calibrate $E_{ph}^0$ as a spectral field-sensor. Peak position is determined by bi-Gaussian peak fits to the modeled $I_{PR}$ spectra shown in Fig. 3c, and the determined $E_{ph}^0$ values are plotted as a function of $F_{max}$ in Fig. 3e. This relation between $E_{ph}^0$ and $F_{max}$ is used as a transfer function to convert the measured peak positions (Fig. 2b) to estimated $F_{max}$ values, plotted as a function of the experimentally applied bias ($V_{exp}$) in Fig. 3f. We see clear evidence of a field-limit from this spectral data; the spectrally determined value of $F_{max}$ does not increase continuously with $V_{exp}$ and reaches a saturated value precisely at the theoretical breakdown limit of β-Ga$_2$O$_3$ i.e. ~6 MV/cm [23,24]. Note that our analysis above made no assumptions about the breakdown limit, and the field values in Fig. 3f are entirely based on the measured photocurrent spectra modeled using the XFK effect and the exciton parameters of β-Ga$_2$O$_3$ previously described. The dark current of the Schottky diode is measured with increasing reverse bias, shown in Fig. 3g. As expected, the current starts to decrease faster with increasing $V_{exp}$ just where the saturation behavior of $F_{max}$ begins in Fig. 3f. The expected reverse breakdown will happen around $40\ V < V_{exp} < 60\ V$) [72].

Because the photocurrent is excited at the focal point of the optical excitation at a particular $(x, y)$ position on the SBD surface, the observed saturation of $F_{max}$ at $V_{exp}$ >30 V is a locally measured phenomenon. In this field range, where local field saturation is observed, near or at the onset of dielectric breakdown, the global device still has not exhibited any current spiking, which occurs in devices processed from this same wafer at $40\ V < V_{exp} < 60\ V$. At the $(x, y)$ position of the $I_{PR}$ measurements, the local $F_{max}$ increases evermore slowly with applied bias in the range 30 V $< V_{exp} <$ 40 V, and therefore $F_{max}$ must necessarily be increasing in other regions of the device that have not yet reached the breakdown limit. Thus, this measurement is sensitive to local field non-uniformity and could be used to map out likely breakdown pathways without destroying devices.

### III.  NON-LINEAR FIELD-DEPENDENT PHOTOCURRENT AMPLITUDE

The amplitude variation observed in Fig. 2g is explained by a field-dependent η, discussed below, due to exciton dissociation and self-trapped hole ionization.

#### A.  Field-dependent photocurrent production pathways: rate equations

The proposed photocurrent generation mechanism is illustrated in Fig. 4, with two possible pathways. Both begin by below bandgap photons absorbed in β-Ga$_2$O$_3$ producing free excitons (X) consisting of a conduction band electron and a valence band hole that are mutually bound by $E_X^0$ =180-270 meV [60,61]. Due to the polaron formation affinity [63] and large effective mass [42,60,61] of holes, the photo-excited holes cause a lattice distortion that induces a short-range trapping potential for the holes, forming a STH within τ$_{ST}$ < 0.5 ps at 295 K [67]. Thus, X that are not dissociated (path 1) become STXs (path 2) consisting of a conduction band electron bound to an STH. The X can also recombine, but this time scale (>1 μs) [66] is much greater than τ$_{ST}$ due to the indirect bandgap, and therefore there is negligible X recombination, a conclusion also supported by the lack of any reports of free-X photoluminescence, i.e. lack of emission peak at ~4.6 eV range in β-Ga$_2$O$_3$ . The fraction of X that dissociate into free carriers along path 1 is, therefore, $\eta_X = D_X/(D_X + \tau_{ST}^{-1})$, where $D_X$ is the field-dependent dissociation rate of X.

In steady-state, charge neutrality requires that photoelectron and photohole currents be identical. Given the larger effective mass of holes than of electrons, we assume the photocurrent is limited by photohole collection. Free holes have an average mobility of 20 cm$^2$/(Vs) as recently reported [79], which is consistent with earlier measurements of long minority hole diffusion lengths of ~200-300 nm [74]. With depletion widths of <1000 nm, the electron and hole drift times are <1 ps (assuming saturation velocities ≥10$^6$ cm/s), while the free carrier recombination times are ~210 ps [74]. Thus, the collection efficiency of free electrons or holes is ~100% over the experimental field range, and the total quantum efficiency of path 1 is $\eta_1 = \eta_X = D_X/(D_X + \tau_{ST}^{-1})$.

The fraction of X that become STX along path 2, is $(1 - \eta_X)$. Within the STX, the large lattice distortion binds the hole by 490 meV [64], localizing it within a single bond distance centered over an oxygen anion, rendering it immobile [16,42,44,63]. As the total STX binding energy is 680 meV [64], within the STX quasiparticle the electron is bound to the STH by (680-490) = 190 meV. As this energy is much smaller than the STH binding energy, the STX dissociation occurs by the electron tunneling into the conduction band at a field-dependent rate of $D_{STX}$. The fraction of STX that are dissociated is, $\eta_{STX} = D_{STX}/(D_{STX} + R_{STX})$, where $R_{STX}$ is the STX recombination rate (~$10^6$ $s^{-1}$) [66]. To produce any current in steady-state, the remaining STH must be field-ionized (490 meV) at a rate of $D_{STH}$ in order to generate mobile free holes at an efficiency of $\eta_{STH} = D_{STH}/(D_{STH} + R_{STH})$. The total fraction of X that become STX and subsequently contribute to current (path 2) is thus, $\eta_2 = (1 - \eta_X) * \eta_{STX} * \eta_{STH}$.

To validate this model, a relationship between $F$ and $\eta$ is derived and compared with the measured $I_{PR}^0$. From the spectral $E_{ph}^0$ data combined with the Stark-modified XFK model, a relation between $V_{exp}$ and $F_{max}$ was previously established, Fig. 3f. Utilizing this relation, the $I_{PR}^0$ data in Fig. 2c are re-plotted as a function of $F_{max}$ in Fig. 5a. A sudden, threshold-like increase of amplitude at about ~4.5 MV/cm is observed. The total IQE is $\eta = \eta_1 + \eta_2$. However, if we assume that $\tau_{ST}$ ~= 0.5 ps [67], then $\eta_1 \gg \eta_2$ over the entire field range and negligible current would be produced along path 2 at all field values. The data in Fig. 5a indicate this is not the case. We hypothesize that the X dissociation (path 1) dominates the photocurrent generation in β-Ga$_2$O$_3$ at lower fields, and the STX dissociation plus STH field-ionization (path 2) must dominate at higher fields i.e. $\eta_2$ turns-on at ~4.5 MV/cm.

**B. Field-dependent tunnel ionization rates of excitons and self-trapped-holes**

The field-induced tunnel ionization rates for excitons ($D_X$), self-trapped excitons ($D_{STX}$), and self-trapped holes ($D_{STH}$) are quantitatively modeled utilizing the field-ionization model of impurities proposed by Chaudhuri *et al.* [80,81]. We note the astonishing similarities of β-Ga$_2$O$_3$ excitons (a light electron bound to a very heavy self-trapped hole) with hydrogenic impurities treated by Chaudhuri *et. al.*'s model. Indeed, the exciton dissociation problem was treated previously with models [46,82,83] that match the hydrogenic exciton picture [84]. The field-dependent tunnel ionization rate is given by,

$$D(F) = \omega \left(\frac{\alpha}{F}\right)^{2n^*-1} e^{-\frac{\alpha}{F}} \quad \ldots (5) \quad ,$$

where $n^*$ is the effective principle quantum number of the ground state within quantum defect theory, and,

$$\alpha = \frac{4(2m^*)^{1/2} E_B^{3/2}}{3q\hbar}$$

$$\omega = \frac{6^{2n^*} E_B |s(n^*)|^2}{3\hbar \Gamma^2 (n^*+1)}$$

$$|s(n^*)| = \frac{\pi}{\sin(n^*\pi)} \left(\frac{1}{2} \sum_{m=0}^{\infty} \frac{1}{(n^*-m-1)^2 (n^*-m)^2}\right)^{-1/2}$$

The reduced/effective mass ($\mu_{ex}$, $m_e^*$, or $m_h^*$) is $m^*$ and F is set equal to $F_{max}$. One should note that, unlike the Merkulov model, the Chaudhuri model incorporates the Stark effect [85]. Thus, incorporating the $bF^2$ term in $E_B$ would be redundant, i.e. $E_B = E_x^0$. The non-integer effective principle quantum number ($n^*$) corrects for deviations from the purely hydrogenic case ($n^* = 1$), and is estimated by [81]:

$$E_B[eV] = \frac{m^* \, 13.6}{m_0 (Kn^*)^2} \ldots (6)$$

For X, with $E_B = E_x^0 = 270 \, meV$ [60], $\mu_{ex} = 0.3 \, m_0$ [60], and $K = K^* = 3.99$ [60], then $n^* = 0.97$. In the case of the STX, replacing $E_B = E_x^0 = 680 - 490 = 180 \, meV$ [65,66], then $n^* = 1.16$. Both values of $n^*$ are fairly minor deviations from ideal hydrogenic states, which is consistent with X and STX Bohr radii ($a_B = 0.70$ nm) that are several bond distances (0.19 nm) [86]. However, when the light electron is no longer present, as in the case of STHs, all that remains is the highly-localized heavy hole expected to exhibit deep-level behavior along with a local lattice distortion (polarization). In this case, an effective dielectric constant ($K_{STH}$) is not known, and neither is the effective mass of the host β-Ga$_2$O$_3$ valence band ($m_h^*$). In previous literature no conclusion could be drawn about its value except that it is >10 $m_0$ [60,61].

There are several recent *ab initio* atomistic models of the STH in β-Ga$_2$O$_3$ that provide a definite length scale for the lattice distortion associated with the STH by calculating the probability density function of the hole, which occupies a p-orbital of an O$^{2-}$ site [87–89]. From these calculations, $a_B$ of the STH is roughly half the average inter-ionic bond distance, indicating a strong departure from the hydrogenic cases of the X and STX. Guided by these first principles calculations, we consider $a_B^* = 0.06 - 0.14$ nm, which limits the range of $K_{STH} * (m_0/m_h^*) = 1.1 - 2.7$. Under this constraint, Eqns. (5-6) can be used to calculate $D_{STH}$ with only one unknown parameter, $m_h^*$. As will be discussed below, good quality fits ($R^2 \geq 0.9$) are obtained for a range of $m_h^* = 18 - 25\ m_0$ for the considered range of $a_B^*$. However, the highest quality fit ($R^2 \sim 0.92$) is obtained for $m_h^* = 18.8\ m_0$, with $a_B^* = 0.127$ nm ($K_{STH} = 45$ and $n^* = 0.51$). The reduced $n^*$ and increased $K$ are consistent with a deep-level in a distorted dielectric environment.

Utilizing Eq. (5) with $m_h^* = 18.8\ m_0$ ($n^* = 0.51$) and $\tau_{ST} = 0.045$ fs, $D_X$, $D_{STX}$, and $D_{STH}$ as a function of $F$ are calculated. The inverses of these rates, i.e. dissociation times, are plotted on a logarithmic scale in Fig. 5b. Clearly $D_{STH}$ (tunnel-ionization of self-trapped holes) is the rate limiting process for photocurrent generation. We obtain quantum efficiencies for both paths ($\eta_1$ and $\eta_2$), plotted in the inset of Fig. 5b. The rise in experimentally measured $I_{PR}^0$ at ~4.5 MV/cm is obtained by path 2's behavior (STX→STH→free carriers). Path 1 (X dissociation) dominates the photocurrent at low field, where the STH produced along path 2 cannot be ionized quickly enough to avoid recombination. But at sufficiently large fields, path 2 can contribute once $D_{STH}$ outpaces $R_{STH}$.

The hole effective mass, $m_h^*$ strongly impacts the tunneling rate of holes out of the self-trapping potential (Eq. (5)), and therefore $m_h^*$ is well determined by the threshold field at which path 2 begins to contribute to photocurrent. This is illustrated in Fig. 5c; keeping $\tau_{ST}$ = 0.045 fs, if $m_h^*$ < 17.5 $m_0$, the path 2 turn-on happens too early (<4 MV/cm), while for $m_h^*$ > 25.5 $m_0$ the turn-on happens too late (>5 MV/cm). The best fit hole mass ($m_h^* = 18.8\ m_0$) matches the observed $I_{PR}^0$ high-field turn-on at ~4.5 MV/cm.

The self-trapping time ($\tau_{ST}$) of holes determines if any photocarriers can follow path 2 (i.e. $\eta_1/\eta_2$), and thus the ratio of the low field (path 1) and high-field (path 2) photocurrent magnitudes. This is illustrated in Fig. 5d; keeping $m_h^*$= 18.8 $m_0$, if $\tau_{ST}$ is <0.01 fs, the low-field photocurrent is too small, while if $\tau_{ST}$ is >1 fs, then no high-field turn-on is observed. We note that the estimated $\tau_{ST}$ = 0.045 fs is consistent with the ultrafast pump-probe absorption data in Ref. [67]. taken at lower photon energies noting the time-resolution limitation of those experiments. Using this field-dependent model of $\eta$, Fig. 5e shows the modeled $I_{PR}$ spectral behavior at varying $F$ values. The calculated $I_{PR}^0$ values are normalized by the highest-field value and plotted as a function of F in Fig. 5f. The model (line) shows a good match to the normalized data (points) taken from Fig. 5a.

## IV. CONCLUSIONS AND OUTLOOK

The optoelectronic properties of ultra-wide bandgap low-symmetry materials, like β-Ga$_2$O$_3$, are far less explored compared with other well-established semiconductors. Using a rather simple measurement (photocurrent spectroscopy in the sub-bandgap regime), a deeper understanding of electron-hole interaction in the strong-field limit was obtained. Just as in 2D TMDs [14,50], the electron-hole interaction in β-Ga$_2$O$_3$ takes on a 2D Hydrogen-like geometry leading to strong polarizabilities combined with large exciton binding energies. The result is that under large electric fields, the electron-hole interaction becomes even stronger, i.e. Stark effect, leading to a prominent redshift of the exciton absorption peak.

The exciton peak position was shown in β-Ga$_2$O$_3$ to serve as an accurate sensor of the electric field maximum in a Schottky diode structure. We observed a local electric field saturation effect, where the local field does not increase past ~6 MV/cm (nearly the breakdown field of β-Ga$_2$O$_3$) even though the experimentally applied bias increases (as does the photocurrent amplitude) indicating that this technique can be used to map out field non-uniformity and explore the onset of dielectric breakdown in a non-destructive manner, before actual device breakdown (current spiking) occurs.

Not only does this measurement provide a simple spectral means to track the local field, but the amplitude of the photocurrent versus electric field reveals a very unusual field-dependent quantum efficiency. At ~4.5 MV/cm, the peak amplitude shows an abrupt turn-on (threshold). This non-linear phenomenon is understood by developing the photocurrent rate equations for β-Ga$_2$O$_3$ considering the free exciton (X), self-trapped exciton (STX), and self-trapped hole (STH) field-dependent dissociation rates and ionization rates. As the excitons and STH in β-Ga$_2$O$_3$ are strongly bound, and the effective masses of holes are large, a deep-tunneling based field-ionization model was employed to model the ionization rates. Using literature values of the carrier lifetimes, the amplitude variation with field was fit to estimate two fundamental properties of the β-Ga$_2$O$_3$ valence band, the hole effective mass (18.8 $m_0$) and the hole self-trapping time (0.045 fs) assuming an STH Bohr radius of 0.127 nm. Although this simplistic isotropic model ignores the anisotropy of the STH state, the results are consistent with previous atomistic *ab initio* studies, indicative of STH localized within $0.06 - 0.14$ nm. Within quantum defect theory, we find a reduced effective principle quantum number $n^* = 0.51$, which is consistent with the STH exhibiting deep-level character and an increased polarizability due to the local ionic distortion.

For a long time, strong-field exciton physics in 3D solids has evaded direct measurement and quantitative modeling. This is partly due to the scarcity of 3D materials with simultaneously large $E_x$ and $a_B$, and partly due to limitations arising from smaller breakdown fields even when such materials can be found. As a result, this regime of exciton field-polarization physics was reported only for 1D quantum wires and more recently in 2D TMD materials [14,50,90]. Even then, the models could only qualitatively explain the results unless complex numerical methods were employed. However, in β-Ga$_2$O$_3$, the exciton Stark and Franz Keldysh phenomena can be explained using a simple modified-Wannier-Mott based model [36,39,81] . With the advent of a non-destructive optoelectronic method to measure electric field based on a spectral peak position, it may become more routine to quantify and study excitonic absorption and dissociation mediated photocurrent generation processes in semiconductors with strongly bound and anisotropic excitons.

**Acknowledgments** Funding for this research was provided by the Center for Emergent Materials: an NSF MRSEC under award number DMR-1420451 and by the AFOSR GAME MURI (Grant FA9550-18-1-0479, Program Manager Dr. Ali Sayir).

**Figure Captions**

Fig. 1: Electric-field modified exciton absorption, dissociation, and carrier ionization processes in the depletion region of a β-Ga₂O₃ Schottky barrier diode. (a) The photon flux is focused to a point $(x, y)$ on the device surface, while an applied reverse bias and Schottky barrier cause a depleted region with a non-zero electric field (F) along the z-axis. Absorbed sub-bandgap photons generate neutral free eXcitons (X) formed by Coulombic electron and hole interaction and different ionization processes take place leading to photocurrent from drifting free carriers. F distorts the conduction band (CB) and valence band (VB) edges which allows e-tunneling based dissociation of X. Within X, holes generate a lattice distortion that locally binds holes generating neutral Self-Trapped eXcitons (STX), which themselves can dissociate by e-tunneling. Afterward, a Self-Trapped Hole (STH) is left behind. For these immobile charges to contribute to current, they must be field-ionized by h-tunneling out of the trapping potential. (b) Band edge diagram of the Schottky diode structure and layer structure. (c) Schematic of absorption coefficient (α) modified by F, the Franz-Keldysh effect. The sub-band-gap X absorption peak is red-shifted with the field, whereas the band edge absorption edge shows a field-induced broadening.

Fig. 2: Polarization and frequency dependent photocurrent spectroscopy of β-Ga₂O₃ Schottky diodes. (a) Schematic of device structure on a β-Ga₂O₃ (010) oriented crystal, and excitation geometry showing the polarization angle and crystallographic axes. (b) Representative photocurrent spectra normalized by excitation power (photoresponsivity, $I_{PR}$) at various values of reverse bias ($V_{exp}$). Black circles indicate the peaks in $I_{PR}$ as fitted by "bi-Gaussian" peak fitting routine. (c) Same data plotted on a semi-logarithmic scale. (d) Polarization-angle (θ) dependence of the photocurrent peak position ($E_{ph}^0$) relative to the peak position at θ = 0° (parallel to [100]) revealing the anisotropy of exciton absorption in β-Ga₂O. Line is a guide to the eye. (e) Peak amplitude ($I_{PR}^0$) as a function of illumination time (plotted on a semi-logarithmic scale). (f) Peak position ($E_{ph}^0$) as a function of $V_{exp}$. (g) Peak amplitude ($I_{PR}^0$) as a function of $V_{exp}$.

Fig. 3: Spectral observation of the breakdown limit of β-Ga₂O₃ by local electric field (F) measurement. Two hypothetical electrostatic conditions are simulated. (a) Field profiles $F(z)$ at with varying peak field values ($F_{max}$), but with applied bias (V) held constant. (b) $F(z)$ with constant $F_{max}$, but varying V. (c) and (d) plot the photoresponsivity spectra modeled using Eqs. (2-4) (assuming η = 1) for the field profiles shown in (a) and (b), respectively. (e) Peak position ($E_{ph}^0$) extracted from the simulated spectra of (c) and plotted as a function of $F_{max}$ demonstrating $E_{ph}^0$ as a spectral detector of $F_{max}$. (f) The data of Fig. 2b are replotted by using (e) as a transfer function to convert the spectrally measured $E_{ph}^0$ values to estimated $F_{max}$ values that are then plotted as a function of the experimentally applied reverse bias ($V_{exp}$). The spectrally estimated $F_{max}$ saturates near the theoretical breakdown field of β-Ga$_2$O$_3$ before global device breakdown is observed. Line is a guide to the eye. (g) The dark I-V of the device shows the beginning of the breakdown, which occurs in between 40 to 60 V.

Fig. 4: Photocurrent production pathways in β-Ga$_2$O$_3$ due to field-induced neutral free-eXciton (X) dissociation and carrier ionization processes. X dissociate at a rate $D_X$ by e-tunneling into the conduction band (CB) generating free electrons and holes that drift to the electrodes along photocurrent path 1. Alternatively, holes can become self-trapped at a rate $\tau_{ST}^{-1}$ forming self-trapped eXcitons (STX), which themselves are dissociated by e-tunneling at a rate $D_{STX}$, otherwise, they recombine at a rate $R_{STX}$. If STX dissociate, they leave behind a Self-Trapped Hole (STH) that itself can either dissociate at a rate $D_{STH}$ by h-tunneling into the valence band (VB), otherwise, they recombine with free-electrons at a rate $R_{STH}$. Path 1 is preferred at the low field with a probability of $\eta_1 = \eta_X$, while path 2 becomes possible at high-field with a probability of $\eta_2 = (1 - \eta_X) * \eta_{STX} * \eta_{STH}$.

Fig. 5: Field-dependent dissociation of self-trapped excitons and holes in β-Ga$_2$O$_3$. (a) Photoresponsivity peak amplitude ($I_{PR}^0$) as a function of field maximum ($F_{max}$) obtained by replotting the data of Fig. 2c and converting the experimental reverse bias to $F_{max}$ using Fig. 3f. (b) Dissociation times as a function of $F$ calculated using Eq. 5. Inset: resulting quantum efficiencies of the two photocurrent production paths (see Fig. 4) as a function of F. The total quantum efficiency, η as a function of field $F$ with varying (c) hole effective mass $m_h^*$) and (d) self-trapping time ($\tau_{ST}$) illustrating the sensitivity of the exciton and self-trapped hole field-

ionization model to these material parameters. (e) Normalized photoresponsivity ($I_{PR}^N$) spectra at various $F$ modeled using Eqs. (2-5), i.e. $\eta(F) \neq 1$. (f) Normalized $I_{PR}^{0N}$ as a function of F, comparing the data and the best fit. Data are from (a) and theory from (e), where the $I_{PR}^0$ are normalized by the maximum value of $I_{PR}^0$.

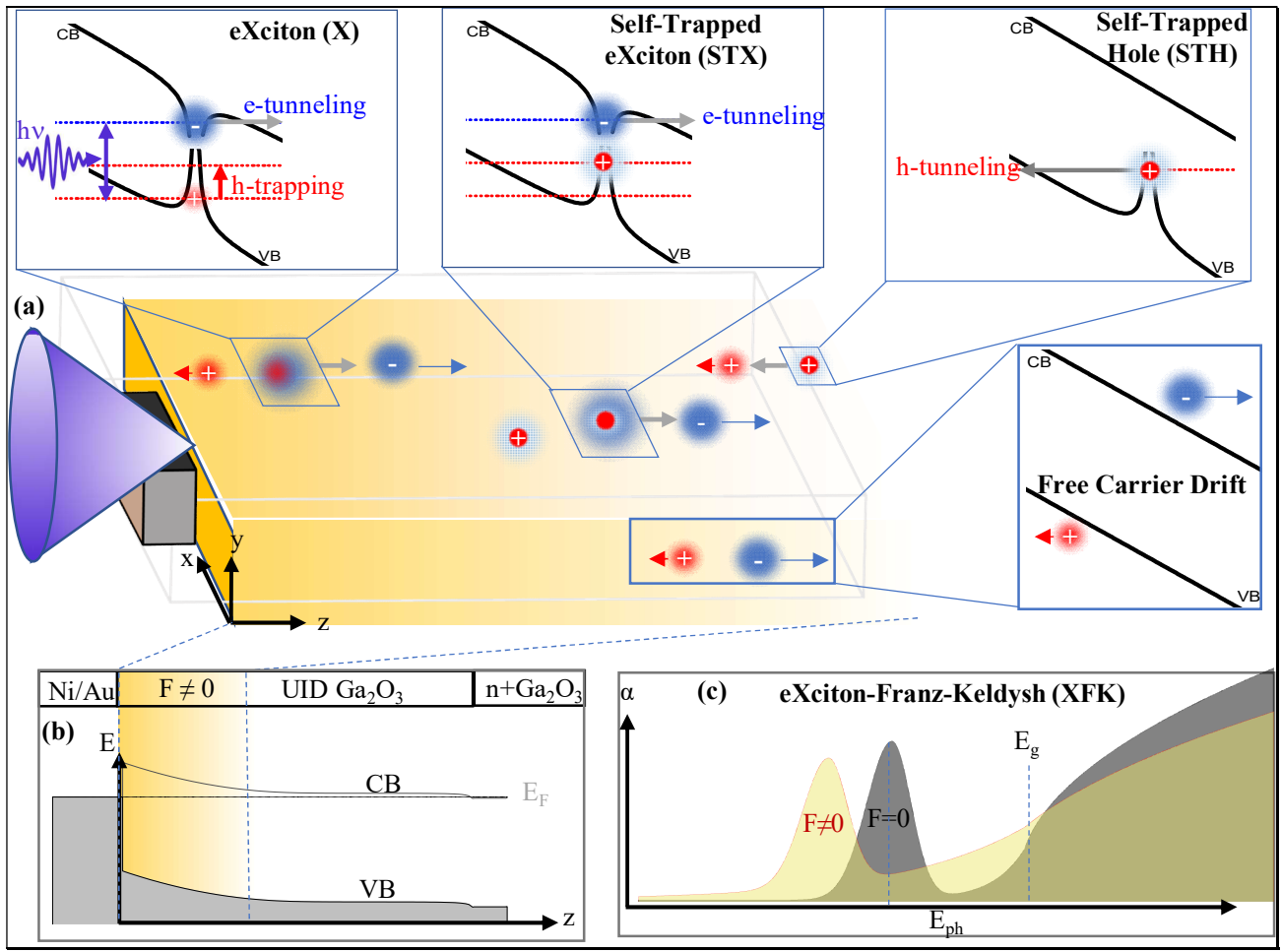

Fig. 1. Adnan *et al.*

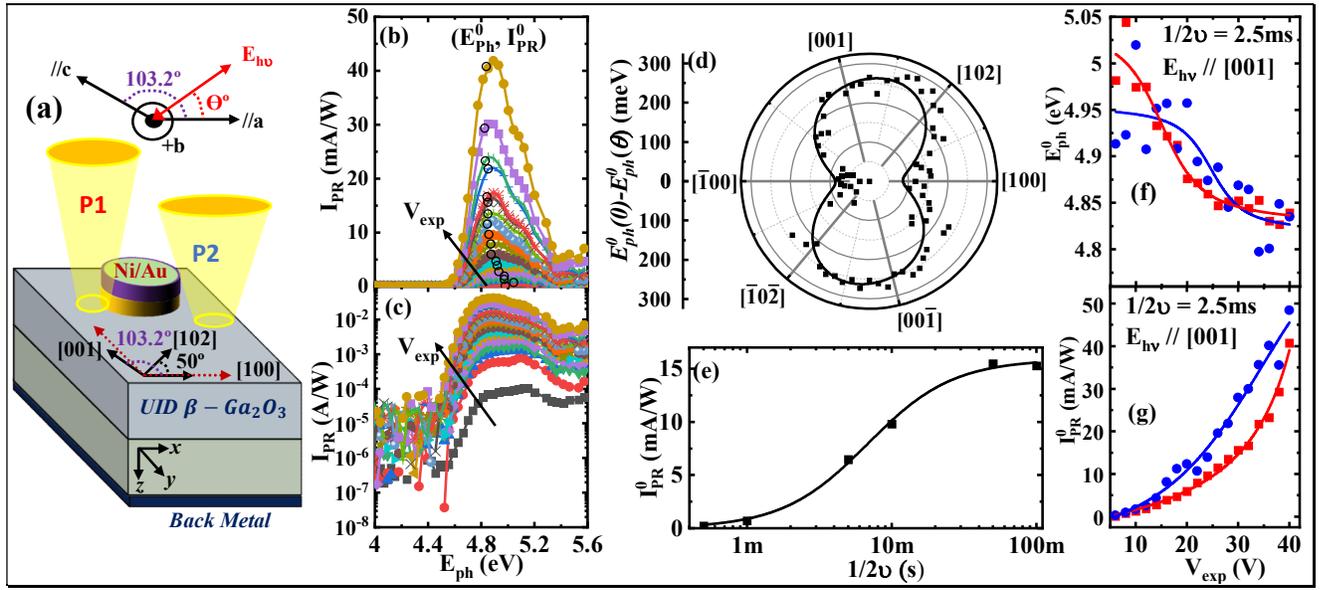

Fig. 2. Adnan *et al.*

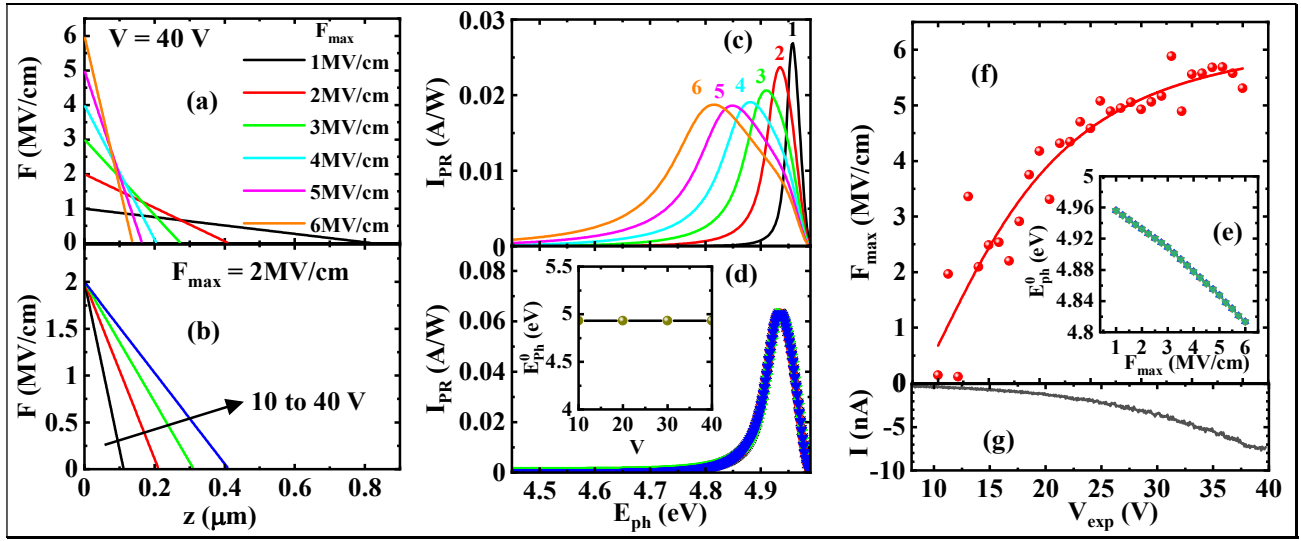

Fig. 3. Adnan *et al.*

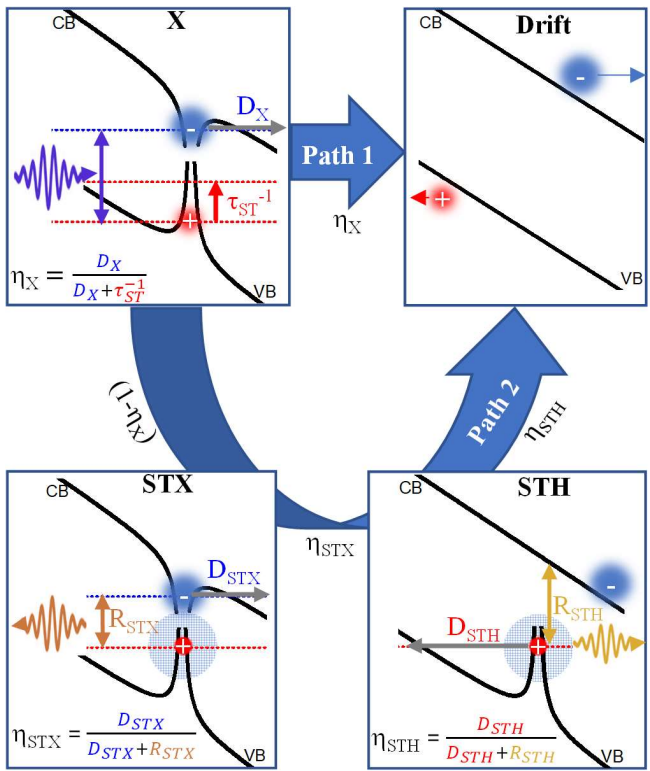

Fig. 4. Adnan *et al.*

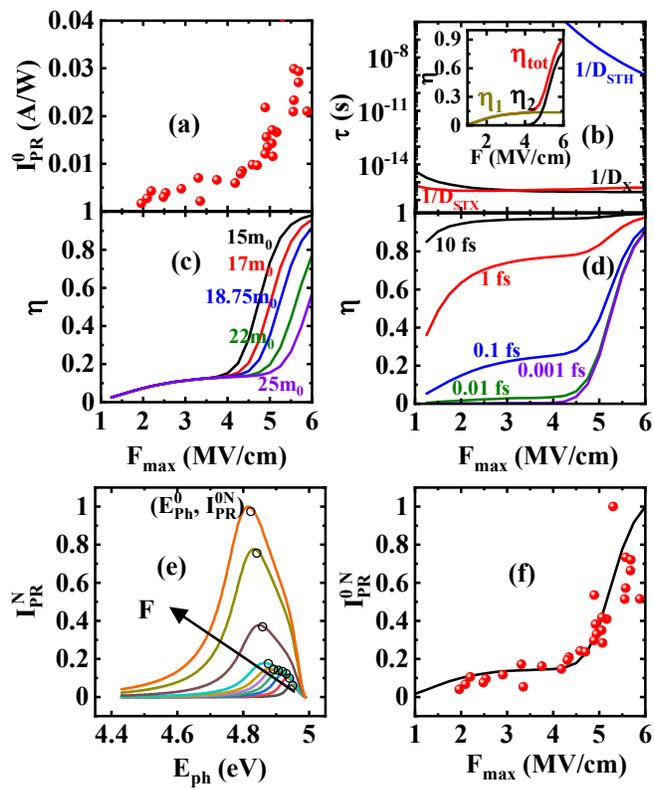

Fig. 5. Adnan *et al.*

# Supplemental Materials: Spectral measurement of the breakdown limit of β-Ga$_2$O$_3$ and field-dependent dissociation of self-trapped excitons and holes

Md Mohsinur Rahman Adnan[1,&], Darpan Verma[2,&], Zhanbo Xia[1], Nidhin Kumar Kalarickal[1], Siddharth Rajan[1], Roberto C. Myers[1,2]

[1]Department of Electrical and Computer Engineering, The Ohio State University, Columbus, Ohio 43210, United States

[2]Department of Material Science and Engineering, The Ohio State University, Columbus, Ohio 43210, United States

## I. *Optical path of photocurrent measurement:*

The cross section of the vertical Schottky diode with different identified regions is shown in Fig. S1. The diode has a top metal contact where the reverse bias is applied with respect to the bottom metal contact. Any position on the surface can be denoted by the co-ordinates $(x, y)$; while the third dimension into the vertical structure is denoted by $z$ axis. As a result, the potential drop and field profile at any given position have their variations along $z$ axis. The representative measurement of responsivity (Fig. 2b) in the main text had been carried out by illuminating a position P1 $(x_1, y_1)$ very close to the top metal contact, while another measurement had been carried out further away from it, at P2 $(x_2, y_2)$ (red vs. blue data of Fig. 2f and 2g). Here, we describe the optical path at P1. The photons enter the diode structure and travel along the z-axis.

---

[&]These authors contributed equally to the work.

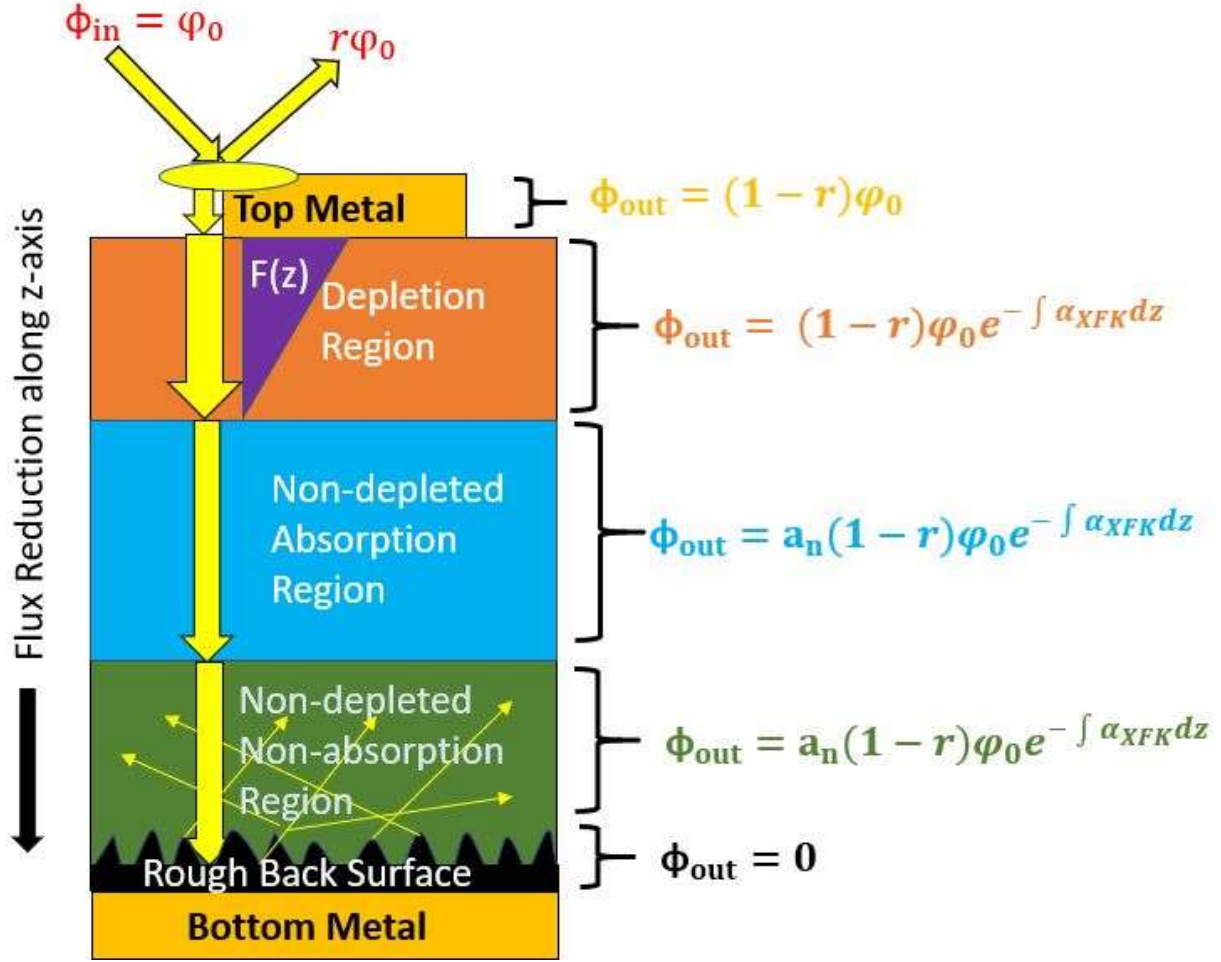

*Figure S1: Cross-section view along z-axis of the Schottky diode.*

The photon flux that reaches the top surface either transmits into the vertical structure or reflects from it. In general, if the amount of flux reaching the surface is $\Phi_0$ and the reflectivity of the top surface is $r$ then the amount of flux that enters the depletion region is,

$$\Phi_{in} = (1-r)\Phi_0 \dots \text{(S1)}$$

Some portion of the excitation light may fall on the top metal and the rest on the semiconductor surface. From a measurement of the reflectivity, shown in Fig. S2, the spectral variation of $r$ over the wavelength range of the measurement below bandgap (3.5-5 eV) is small. β-$Ga_2O_3$ and Ni/Au both show low reflectance across this range around 15% reflectance varying only by +/- 4%. This suggests that from the reflectivity viewpoint we need not distinguish between the metal contact and semiconductor surface. But from the transmission viewpoint, the flux that enters the vertical structure is different depending on the illumination position (top metal/semiconductor surface). The Ni+Au metal contact has a thickness of 130 nm. The transmittance of UV light through such a structure will be very small (T<0.00008 at these wavelengths). Thus, photon flux that enters the vertical structure must be from the illumination that impinges upon the semiconductor surface, as shown in Fig. S1.

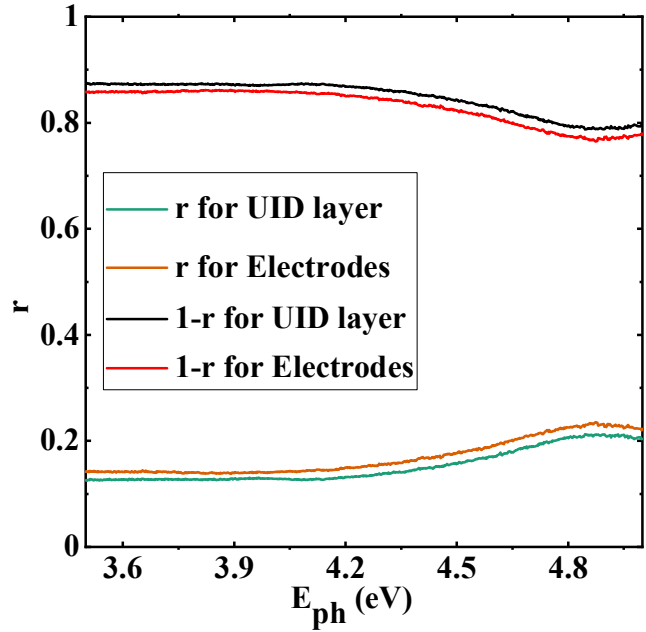

*Figure S2: r and (1-r) plotted over the wavelength range below bandgap (5eV).*

As photon flux $\Phi_{in}$ enters the vertical structure, it will be traversing the depletion region just below the metal/ semiconductor interface i.e. top junction. The field profile $F(z)$ in the depletion region of a Schottky junction has a triangular shape given by equation (4) in the main text and as shown in Fig. S1. This field profile will give rise to the absorption coefficient $\alpha_{XFK}(F(z),\omega)$ according to the Merkulov XFK model described by equation (3) in the main text. As a result, there will be sub-band gap absorption of photon flux in this region. The outgoing flux from the depletion region will have a value,

$$\Phi_{out,d} = \Phi_{in} e^{-\int_0^W \alpha_{XFK}(F(z),\omega)dz} = (1-r)\Phi_0 \, e^{-\int_0^W \alpha_{XFK}(F(z),\omega)dz} \ldots (S2),$$

where W is the depletion width, which is smaller than the UID layer thickness of 1µm at all values of reverse bias.

This outgoing flux from the depletion region will now enter the non-depleted absorption region of the vertical structure. Since the non-depleted layer has negligible electric field, photocarriers could possibly be produced and collected within the minority hole diffusion length, $L_n$, with an absorption factor of $a_n = e^{-\alpha_n L_n}$, where $\alpha_n$ is the zero-field absorption coefficient of n-doped β-Ga$_2$O$_3$. Thus, the flux that leaves the possible photocurrent producing regions will be,

$$\Phi_{out,n-d} = \Phi_{out,d} e^{-\alpha_n L_n} = (1-r)\Phi_0 \, e^{-\int_0^W \alpha_{XFK}(F(z),\omega)dz} e^{-\alpha_n L_n}$$

... (S3)

The sub-band gap photon flux that transmits the entire substrate and reaches the bottom metal contact layer (rough back surface) is equal to,

$$\Phi_{out,bott} = (1-r)\,\Phi_0\, e^{-\int_0^W \alpha_{XFK}(F(z),\omega)dz} e^{-\alpha_n L_n} * e^{-\alpha_n L} \ \text{... (S4)}$$

The n-type β-Ga$_2$O$_3$ substrate is relatively thick (L~0.5 mm); thus the sub-bandgap photon flux incident on the back surface is reduced to $\sim e^{-\alpha_n L} > e^{-1000*.05}$ (see main text), which together with the diffuse hemispherical reflection off the back surface as shown in fig. S1 of the substrate allows us to ignore absorption due to a second optical pass through the depletion region, i.e. $\Phi_{out,bott} \sim 0$ from the back surface of the Schottky diode.

Thus, the flux that was converted into collected photocarriers equals,

$$\Phi_{PC} = \Phi_{in} - \Phi_{out,n-d} = (1-r)\,\Phi_0\, \{1 - e^{-\int_0^W \alpha_{XFK}(F(z),\omega)dz} e^{-\alpha_n L_n}\} \ \text{... (S5)}$$

Assuming all of the absorbed flux is converted into current, i.e. each photon generates an electron hole pair, the photocurrent of the Schottky diode is,

$$I_{PC}(no\ loss) = q\Phi_{absorbed} = q(1-r)\Phi_0\,\{1 - e^{-\int_0^W \alpha_{XFK}(F(z),\omega)dz} e^{-\alpha_n L_n}\} \text{... (S6)}$$

Correcting for recombination losses, we include the internal quantum efficiency, η,

$$I_{PC}(actual) = q\eta\Phi_{absorbed} = q\eta(1-r)\Phi_0\,\{1 - e^{-\int_0^W \alpha_{XFK}(F(z),\omega)dz} e^{-\alpha_n L_n}\} \text{... (S7)}$$

This is the full form of photocurrent generated in the Schottky diode. The term $e^{-\alpha_n L_n}$ had been shown to be ~1.00 for our structure in the main text. Which gives the effective photocurrent equation, i.e. equation 1 in main text.

$$I_{PC} = q\eta \Phi_{absorbed} = q\eta(1-r)\Phi_0 \left\{1 - e^{-\int_0^W \alpha_{XFK}(F(z),\omega)dz}\right\}\ldots \text{(S8)}$$

## II. *Light I-V characteristics of the Schottky diode:*

Fig. S3(a) shows the forward current under both dark and light condition, while S3(b) shows the same with applied reverse bias.

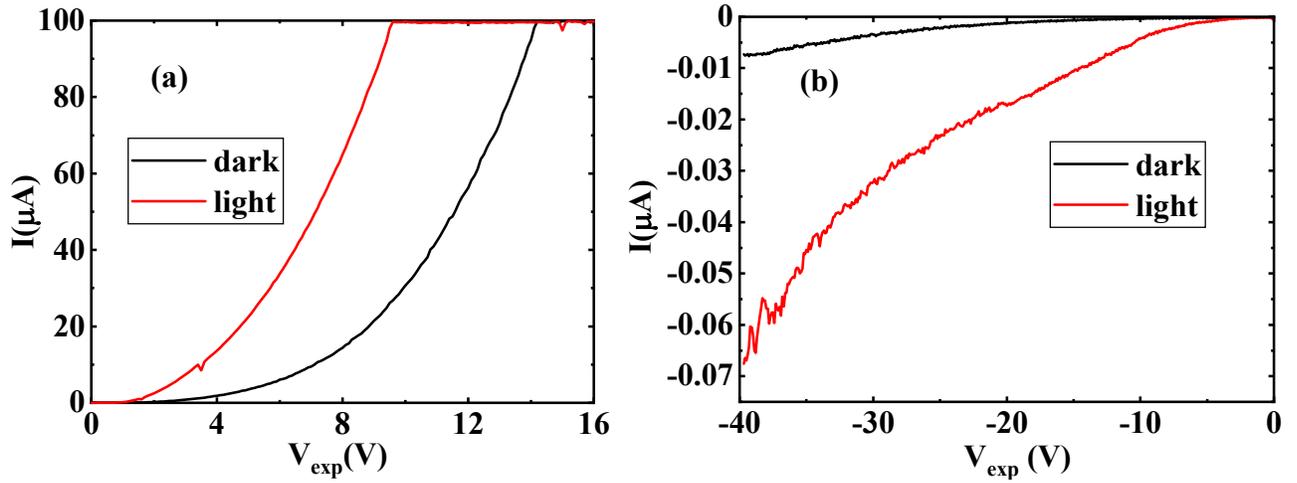

*Figure S3: Forward and reverse current under both dark and light conditions.*

Clearly, illumination plays a vital role in deciding the amount of current measured from the Schottky diode with both forward and reverse bias. The illumination also changes the "turn on" voltage of the diode in forward bias condition. As seen from S3(a), this voltage shifts towards the positive axis by more than 2.5 V without illumination. While the dark condition reverse bias current does not show any sudden large variation near the maximum reverse bias applied (-40 V), the light condition reverse bias current starts to show large variations (with spiking) in the measured current.

## III. *Reproducibility of the XFK effect in β-$Ga_2O_3$ and spectral $F_{max}$ measurement:*

The $F_{max}$ data in Figs. 2, 3 and 5 of the main text were acquired with the polarizer aligned to the c-axis [001] in-plane direction of the (010) oriented crystal. Previous non-polarization resolved measurements were acquired from multiple devices fabricated from $Ga_2O_3$ Schottky devices with a doping thickness of ~1 µm and breakdown voltage around -50 V. In Figure S4, S5 and S6 we plot a representative raw photocurrent spectrum obtained on such $Ga_2O_3$ Schottky diodes in the earlier unpolarized experiments. The data were fitted to equations (2-4) of the main text using $F_{max}$ and bandgap $E_g$ as fitting parameters. The average apparent bandgap ($E_g = 4.9\ eV$) was seen to vary due to uncontrolled polarization. Nevertheless, we were still able to model an $F_{max}$ dependent peak shift due to the XFK effect and exciton Stark effect. Using the same method as in the main text, we mapped the experimental photocurrent peak positions to obtain the $F_{max}$ vs. $V_{exp}$ dependence of the devices. Here, the saturation of $F_{max}$ near breakdown can be observed clearly (Fig. S5). Moreover, by fitting the peak amplitudes with the model described in main text we see the same field dependent dissociation of self-trapped exciton behavior (Fig. S6). The self-trapping time of exciton, $\tau_{STX}$ was chosen as 0.125 fs, within the limit of the range of the self-trapping time by our estimates. The hole mass was kept at $18.8\ m_0$, just as it is in the main text.

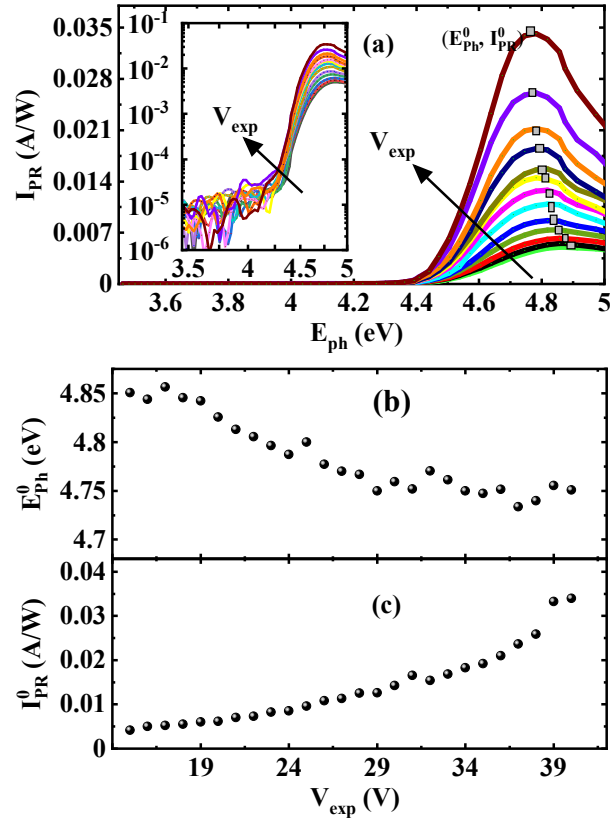

Figure S4: Measured photocurrent spectra on a different Schottky diode without polarization control. (a) The actual spectra (inset in log scale), (b) the peak position and (c) the peak amplitude.

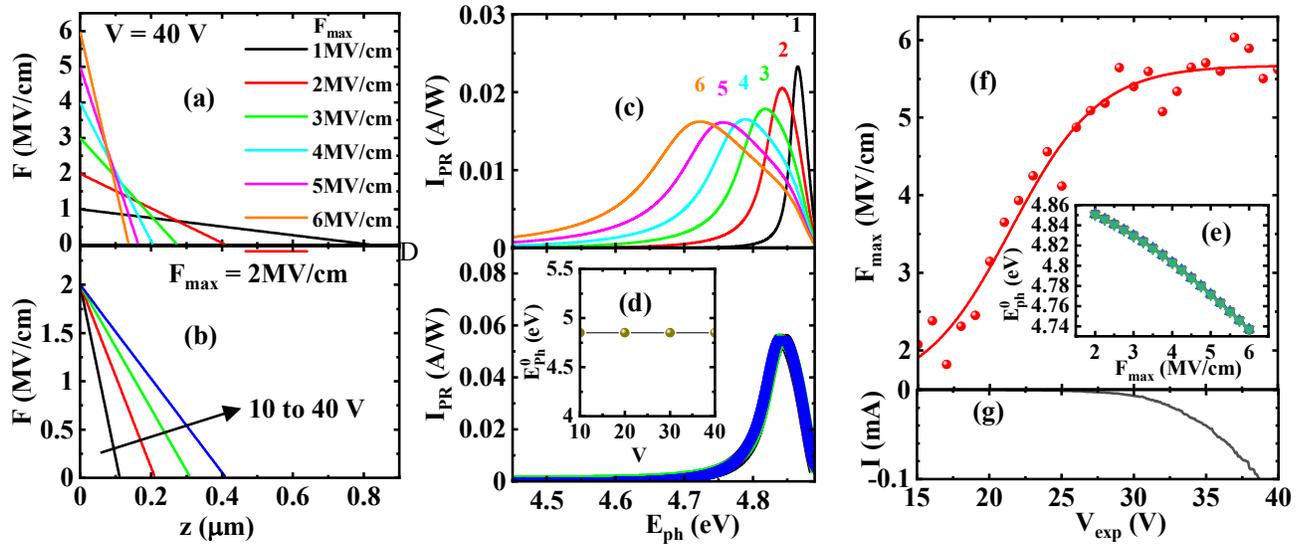

*Figure S5: Similar analysis of the main text figure 3 on the data provided in S4(b). The saturation of $F_{max}$ vs $V_{exp}$ curve around ~6 MV/cm in S5(f) is clearly visible, as is the reverse I-V breakdown of S5(g).*

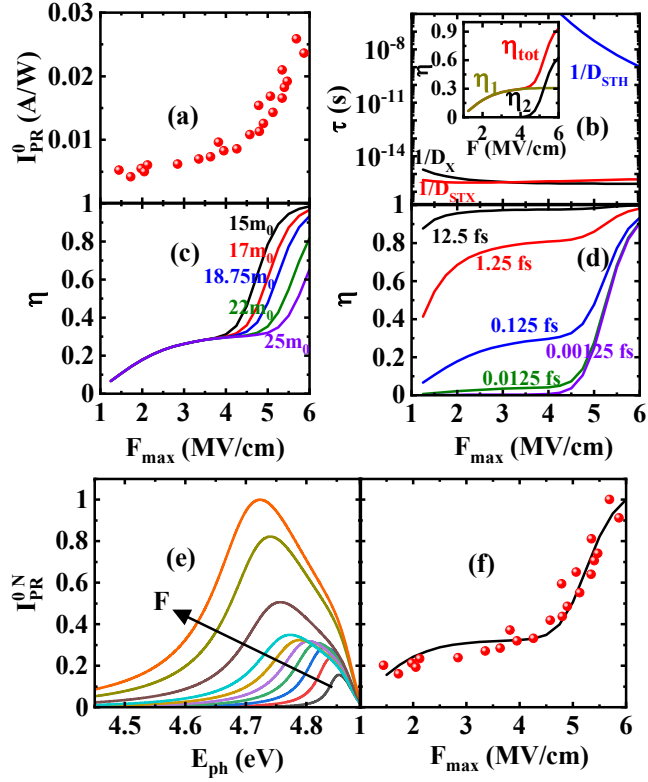

*Figure S6: Similar analysis of the main text figure 5 on the data provided in S4(c). The dissociation of self-trapped exciton around $F_{max}$ ~4.5 MV/cm in S6(f) is clearly visible.*

## IV. *Modulation frequency dependent photocurrent spectroscopy:*

The frequency dependent variation of the maximum amplitude of the photocurrent spectra can be observed in Fig. S7. The measurement was carried out at -30 V bias.

Ref. [28] of main text explains how the chopping frequency ($\nu$), (and the illumination time $\frac{1}{2\nu}$ (s)) impacts the photocurrent in Metal-Semiconductor-Metal (MSM) based $\beta$-$Ga_2O_3$ photodetectors. The structure used in their experiments also utilized Ni/Au top metal on $\beta$-$Ga_2O_3$ thin film structure (Fig. 1 of ref. [28]), similar to our Schottky diodes. The proposed mechanism for photocurrent gain is illustrated in Fig. 5 of ref. [28]. The e-h pairs produced in the $\beta$-$Ga_2O_3$ thin film dissociate with mobile electrons, and less mobile holes. In steady-state the electron and hole currents must be equal by charge neutrality, however time-dependent (transient) phenomena can alter the interface Schottky barrier. As electrons are swept away be the applied electric field, but self-trapped holes remain inside the depletion region, they effectively lower the barrier height at the junction. To maintain charge neutrality, more electrons are needed which are supplied from the metal side. As the Schottky barrier is lowered, the measured photocurrent increases.

This poses a complexity in measuring the photocurrent. Ref. [28] divided the measured photocurrent $I_{ph}^m$ into two parts, the intrinsic bias independent photocurrent $I_{ph}^0$ and the component arising from the barrier lowering $\Delta\varphi_B$, which is proportional to the dark current.

$$I_{ph}^m = I_{ph}^0 + (e^{\frac{\Delta\varphi_B}{KT}} - 1) I_{dark} \qquad \ldots \qquad (S9)$$

Previously [28-29,72] this phenomenon was claimed to explain anomalously large responsivity of β-Ga$_2$O$_3$ MSM photodetectors. Ref. [28] claims the barrier lowering effect to be as large as ~0.3 eV with an applied bias range of only 0-20 V. But this poses a problem for our experiments, since the modeled photoresponsivity spectra would then exhibit bias dependent amplitudes unrelated to exciton dynamics.

Since the problem arises from a slow accumulation of interface charges, which lower the barrier, it can be eliminated from our measurements by lock-in detection in which the optical excitation is chopped (oscillated) at a rate that is faster than the slow charging dynamics. The experiments in Ref. [28] were carried out at a frequency of 30 Hz. At higher frequency chopping, the illumination time is smaller, and the effect of interface charges will be filtered from the lock-in detected photocurrent. Indeed, our measurements of photoresponsivity spectra (Fig. S7 and main text Fig. 2e) with varying frequency show that as the chopping frequency is increased, the responsivity peak becomes smaller. Since our experiments are carried out at a much higher frequency of 200 Hz, we isolate photocurrent that is due only to collected photocarriers and not due to the interface charging phenomena. Increasing the chopping frequency further can provide more certainty in this regard, as shown in Fig. S7.

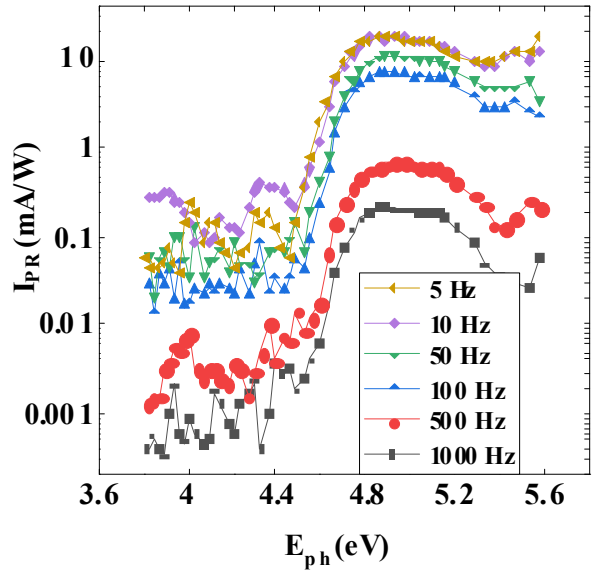

*Figure S7: The frequency dependence of the responsivity spectra of the diode.*

V. <u>*Polarization dependent photocurrent spectroscopy:*</u>

Examples of the raw polarization-dependent photocurrent spectra are plotted at 25° intervals in Fig. S8. The extracted peak positions (at 5° steps) are shown in the polar plot of Fig. 2d of the main text. The movement of peak position is evident from these figures.

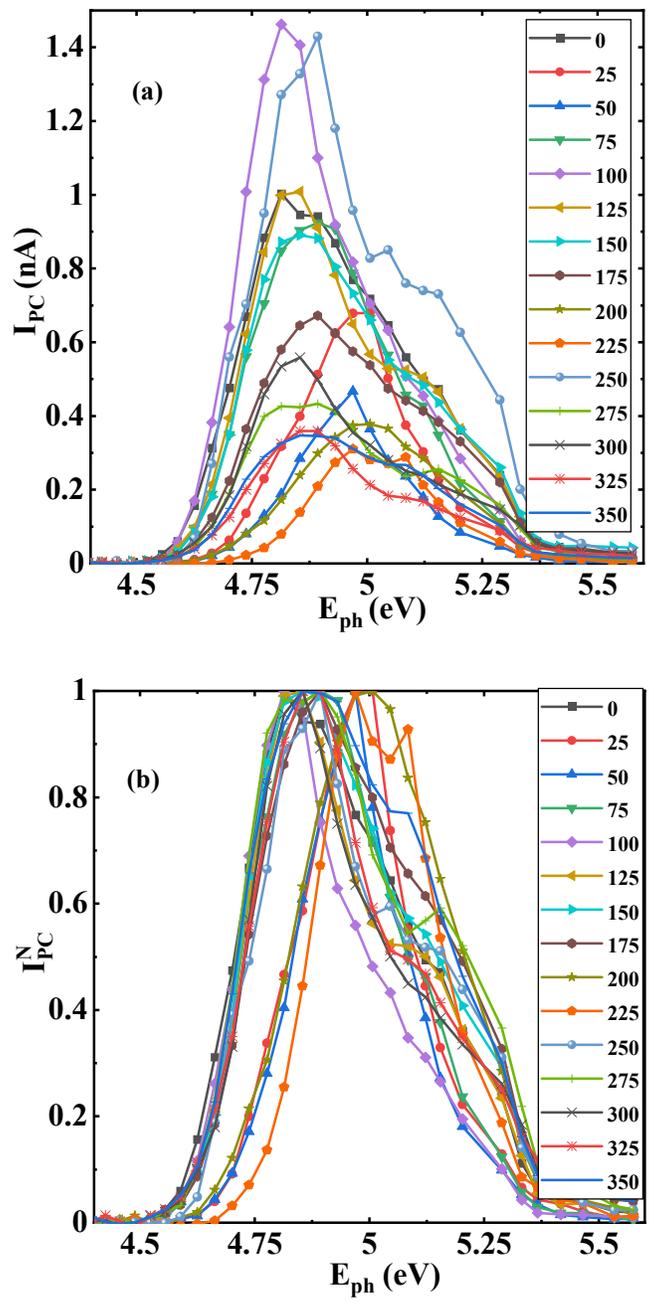

*Figure S8: The angle dependence of the (a) raw (b) normalized photocurrent spectra (experimentally measured).*